\documentclass[fleqn,usenatbib]{mnras}

\usepackage{newtxtext,newtxmath}

\usepackage[T1]{fontenc}

\DeclareRobustCommand{\VAN}[3]{#2}
\let\VANthebibliography\thebibliography
\def\thebibliography{\DeclareRobustCommand{\VAN}[3]{##3}\VANthebibliography}

\usepackage{graphicx}	
\usepackage{amsmath}	
\usepackage[english]{babel}

\DeclareRobustCommand{\VAN}[3]{#2} 

\title[Evaporation from magnetar's surface]{Radiatively driven evaporation from magnetar's surface}

\author[Demidov, Lyubarsky]{
Ivan Demidov\thanks{E-mail: dvsmallville@gmail.com}, Yuri Lyubarsky
\\
Physics Department, Ben-Gurion University, PO Box 653, Beer-Sheva 84105, Israel\\
}

\date{Accepted XXX. Received YYY; in original form ZZZ}

\pubyear{2022}

\begin{document}
\label{firstpage}
\pagerange{\pageref{firstpage}--\pageref{lastpage}}
\maketitle

\begin{abstract}
The luminosity of the Soft Gamma Repeater (SGR) flares significantly exceeds the Eddington luminosity. This is because they emit mainly in the E-mode, for which the radiative cross-sections are strongly suppressed. The energy is released in the magnetosphere forming a magnetically trapped pair fireball, and the surface of the star is illuminated by the powerful radiation from the fireball. We study the ablation of the matter  from the surface by this radiation.
The E-mode photons are scattered within the surface layer, partly being converted into O-photons, whose scattering cross-section is of the order of the Thomson cross-section.  The high radiation pressure of the O-mode radiation expels the plasma upwards.
The uplifted matter forms a thick baryon sheath around the fireball. If an illuminated fraction of the star's surface includes the polar cap, a heavy, mildly relativistic baryonic wind is formed.  
\end{abstract}



\section{Introduction}

Magnetars are neutron stars with the surface magnetic field of the order of $10^{14}-10^{15}$ G \citep{DuncanThompson1992,ThompsonDuncan1995}. The energy of this field feeds all their activity. 
In particular, they sporadically emit X-ray bursts,
which are produced by a sudden restructuring of the magnetic field. There are three types of bursts, depending on their energy release: short bursts ($<10^{41}$ erg), intermediate bursts ($10^{41-43}$ erg), and giant flares ($10^{44-46}$ erg) (see the reviews by \citealt{Turolla2015} and \citealt{Kaspi_Beloborodov17}). 
In giant flares, the released energy could not be confined by the magnetospheric field. Therefore a powerful outflow is formed, producing the so-called hard spike in the light curve with the duration of $\sim 0.5$ s. The rest of the released energy of giant flares and the whole energy of intermediate and weak flares remains confined within the magnetosphere in the form of a radiatively cooling fireball. 

The luminosity of fireballs in giant flares and intermediate bursts, as well as in the strongest short bursts, exceeds the Eddington luminosity. In the strong magnetic field, the radiation propagates in two orthogonally polarized modes having drastically different scattering and absorption cross-sections if the radiation frequency is well below the cyclotron frequency (e.g., \citealt{Meszaros1992}), which is typically the case in magnetar's magnetospheres. The so-called E-mode is polarized perpendicularly to the background magnetic field and therefore only very weakly interacts with matter. For this radiation, the Eddington luminosity is a few orders of magnitude larger than the regular Eddington luminosity \citep{Paczynski86}.  
The fireball in the magnetosphere cools by emitting predominantly in the E-mode. This radiation is thermal with temperatures 10-30 keV. 
\citet{ThompsonDuncan1995} suggested that if the energy flux of E-mode from fireball exceeds the magnetic Eddington flux, $F_\text{edd}^B$, then the radiation obliquely incident on the magnetar ablates material from the surface. 
They claimed that 
the photon splitting produces O-photons, 
which scatter many times and provide a strong radiative force. This ablated material forms a baryon-loaded sheath around the fireball, and the outgoing radiative flux across the magnetic field lines is reduced to $\sim F_\text{edd}^B$. This mechanism of material ablation differs from the another one, also proposed by \citet{ThompsonDuncan1995}, associated with fireball contraction. The magnetar crust below the hot fireball 
absorbs a huge amount of energy and when the fireball above the given area of the surface has evaporated, this energy is released by blowing material off the surface. We do not consider such an ablation mechanism  in our paper, but focus only on the first one. 

The picture of the radiatively driven evaporation from the illuminated surface looks paradoxical because in this case, the radiation falls on the surface from above.
Photon splitting could not contribute to the material ablation if the material was not initially suspended in the magnetosphere because the additional pressure of O-photons prevents ablation. In this paper, we consider in detail the mechanism of material ablation from the magnetar's surface. We argue that the considered ablation mechanism is possible when almost all radiation energy is transported towards the surface in the E-mode. These photons are partly converted into the O-mode photons via scattering on electrons in a narrow surface layer. The resulting super-Eddington outward flux of the O-mode radiation expels the plasma upwards. The photon splitting on the way to the surface prevents the ablation therefore the matter is evaporated only in the vicinity of the fireball, where most of the E-mode photons have not split yet. We will obtain the conditions under which this type of ablation is possible and qualitatively describe the formation of the baryon-loaded sheath around the fireball. If the illuminated region includes the magnetic pole of the star, the ablated material is ejected from the magnetosphere forming a baryonic wind. We will estimate parameters of this wind. 

The paper is organized as follows. In Section 2, we determine the conditions under which incident radiation can lead to material ablation and describe how the baryon-loaded sheath is formed. In Section 3, 
we estimate the mass flow of the evaporated plasma. In Section 4, we describe the properties of the baryon wind thus formed. The conclusions are presented in Section 5.

\section{Conditions for material ablation}


\subsection{Radiation in the magnetar magnetosphere}

In a magnetar flare, the optically thick electron-positron fireball is formed, which is confined by the magnetar magnetic field \citep{ThompsonDuncan1995}. The fireball slowly cools by emitting thermal radiation with the temperature $kT\sim 10-30$ keV. 
In the very strong magnetic field of the magnetar magnetosphere, the vacuum becomes a birefringent medium, so radiation propagates in the form of two normal modes: ordinary (O-mode), which interacts strongly with matter, and extraordinary (E-mode), which interacts weakly with matter. The O-mode radiation is trapped within the fireball therefore only the E-mode photons are emitted away.

At the photon energies and the plasma densities relevant for our study, the main contribution to the photon transparency comes from Thomson scattering on free electrons, the contribution of bound-bound and bound-free transitions being negligible. However, the free-free processes could be important deep inside the surface layer where the temperature is higher. The role of the free-free transitions in the radiative transfer equation is discussed in Appendix B. An important point is that in the scattering process, the modes can convert one into another.  The cyclotron frequency of ions is $\omega_{ci}= ZeB/(Am_pc)$, so that for $B\sim 10^{15}$ G  and 
$A/Z\approx 2$, the cyclotron energy is roughly $\hbar\omega_{ci}\sim 3$ keV. Most of the radiation energy in the case of interest is at higher photon energies, 
therefore, the scattering of photons on ions can be neglected. 

The electron cyclotron energy is well above the photon energy; then 
the differential cross sections for the scattering of a photon with the frequency  $\omega$ may be written as (e.g., \citealt{Meszaros1992})
\begin{equation}\label{00-2}
\begin{split}
&\text{d}\sigma_{\text{E}\rightarrow\text{E}}=\frac{3}{8}\sigma_T\left(\frac{\omega}{\omega_{ce}}\right)^2\text{d}(\cos\theta'),\\
&\text{d}\sigma_{\text{E}\rightarrow\text{O}}=\frac{3}{8}\sigma_T\left(\frac{\omega}{\omega_{ce}}\right)^2\cos^2\theta'\text{d}(\cos\theta'), \\
&\text{d}\sigma_{\text{O}\rightarrow\text{E}}=\frac{3}{8}\sigma_T\left(\frac{\omega}{\omega_{ce}}\right)^2\cos^2\theta\text{d}(\cos\theta'), \\
&\text{d}\sigma_{\text{O}\rightarrow\text{O}}=\frac{3}{4}\sigma_T\!\Big[\sin^2\!\theta\sin^2\!\theta'+\\
&\quad\quad\quad\,\,+\frac{1}{2}\left(\frac{\omega}{\omega_{ce}}\right)^2\!\!\cos^2\theta\cos^2\!\theta'\Big]\text{d}(\cos\theta'), 
\end{split}
\end{equation}
where $\theta$ and $\theta'$ are the angles between the direction of propagation of the photon and the magnetic field before and after the scattering, correspondingly,
$\sigma_T$
is the Thomson cross section,
\begin{equation}
    \omega_{ce}=\frac{eB}{m_ec}=\frac B{B_{\rm QED}}\frac{m_ec^2}{\hbar},
\end{equation}
where $B_{\rm QED}=m_e^2c^3/e\hbar=4.414\times 10^{13}$ G is the critical QED field strength.

These expressions are violated  near the vacuum resonance, where the vacuum and the plasma equally contribute to the wave dispersion; 
this situation is considered in Appendix A. In the strong magnetic field, the E-photon could split into two O-photons \citep{Adler1971}; the role of this process will be considered in Section 2.3.

One sees that in the magnetar's magnetic field, the cross-sections for the E-mode are smaller than the cross sections for the O-mode by a very large factor. This suppression of the E-mode's scattering cross-section is due to the fact that the energy of the first Landau excitation $\sim (2B/B_\text{QED})^{1/2}m_e c^2$ is much higher than the temperature of radiation. Then the emission from the optically thick medium is dominated by the E-mode. In this case, the Rosseland mean cross-section is found as (\citealt{Silantev1980}; \citealt{ThompsonDuncan1995}) \begin{equation}
\sigma_R(B,T)=
5\pi^2\sigma_T\left(\frac{kT}{\hbar\omega_{ce}}\right)^2.
\end{equation}
The corresponding magnetically modified Eddington flux,   
\begin{equation}\label{edd}
F_\text{edd}^B=\frac{m_pgc}{Y_e\sigma_R(B,T)},
\end{equation}
exceeds the standard Eddington flux by orders of magnitude.
Here $g$ is the gravitational acceleration at the surface of the star, $Y_e$ -- the electron number per baryon. 

The temperature of the photosphere emitting the flux $F_\text{edd}^B$ is found  by using the Stefan-Boltzmann law, $F_\text{edd}^B=(1/2)\sigma_{\text{SB}}T^4$, where $\sigma_\text{SB}$ is Stefan – Boltzmann constant, which yields
\begin{equation}\label{edd1}
   kT= \frac{15.7}{Y_e^{1/6}}\left(\frac{B(R)}{10B_\text{QED}}\right)^{1/3}\!\! \left(\frac{g(R)}{2\times10^{14}\,\rm cm\cdot s^{-2}}\right)^{1/6} \,\,\, \text{keV}
    \end{equation}
Then one finds finally
\begin{equation}\label{edd2}
   F_\text{edd}^B = \frac{3.1\times10^{28}}{Y_e^{2/3}}\left(\frac{B(R)}{10B_\text{QED}}\right)^{4/3}\!\! \left(\frac{g(R)}{2\times10^{14}\,\rm cm\cdot s^{-2}}\right)^{2/3}\rm erg/s.
    \end{equation}

\subsection{Conditions for radiatively driven plasma evaporation}




\begin{figure}
	\includegraphics[width=\columnwidth]{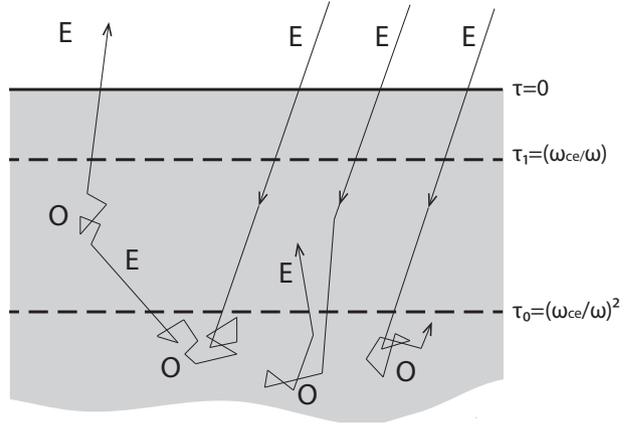}
    \caption{Propagation of E-photons inside star surface layers}
    \label{propagation}
\end{figure}

In this subsection, we find conditions under which a static atmosphere could not exist if it is illuminated from above. 
The fireball emits only in the E-mode. Let us for a while neglect the possible photon splitting; then the surface is hit only by E-mode photons, which penetrate 
the depth $\Delta l_\text{E}\sim 1/\sigma_{\text{E}}N_e$, where $N_e$ is the characteristic electron number density in the considered layer, $\sigma_{\text{E}}=\sigma_{\text{EE}}+\sigma_{\text{EO}}$.  Inasmuch as the emission from the fireball is thermal with the temperature $kT\sim 10-30\, {\rm keV}\ll\hbar\omega_{ce}$, one finds that the Thomson optical depth of this E-mode photosphere is very large
\begin{equation}
    \tau_0=\sigma_TN_e\Delta l_{\text E}\sim \left(\frac{\hbar\omega_{ce}}{3kT}\right)^2.
\label{Ephotosphere}\end{equation}
The E-mode photons are either reflected back after a few scattering or converted into O-photons (see Fig. \ref{propagation}). The scattering cross-sections for O-photons are of the order of the Thomson cross-section.  Therefore the number of scatterings required for O-photons to escape is about $\tau_0^2$ 
whereas the O-photon is converted into the E-mode after only $\sim(\omega_{ce}/\omega)^2\sim\tau_0$ scatterings.  
Therefore O-photons are more likely to convert into E-mode photons than escape. In this and deeper layers, 
the total energy density distributes equally over the two modes, due to mutual transformation of modes into each other and isotropization. Indeed, since the optical depth for the O$\rightarrow$O scattering is large in all directions with a possible exception of a small range of angles $\theta<\tau^{-1/2}$, most photons produced  E$\rightarrow$O transitions have enough time to isotropize. 
Only O-photons produced at the Thomson optical depth $\tau_1\sim\omega_{ce}/\omega$ escape before being converted back into E-mode. This means that most of the incident E-mode radiation is diffusively reflected back from the depth $\sim \tau_0$. A fraction of the incident energy $\sim \tau_1/\tau_0\sim 3kT/\hbar\omega_{ce}\ll\tau_0$ is radiated away in the O-mode polarization. Taking into account that the scattering cross-section of these photons is large, the upward O-mode flux could produce the upward radiation force exceeding the gravity force, which makes the assumed static atmosphere impossible. 


In order to reach quantitative conclusions, we solve the diffusion equation for the O-photon energy density, $\mathcal{E}_\text{O}$, taking into account the mutual conversions of E- and O-photons (see Appendix B):
\begin{equation}\label{00-00}
-\frac{\text{d}}{\text{d}z}\left(D_\text{O}\frac{\text{d}\mathcal{E}_\text{O}}{\text{d}z}\right)=c\sigma_T N_e\alpha^2\left(\mathcal{E}_\text{E}-\mathcal{E}_\text{O}\right).
\end{equation}
Here $z$ is the vertical coordinate, $D_\text{O}=\xi c/\sigma_T N_e$ is the diffusion coefficient along the magnetic field, $\xi \approx 5$, and 
\begin{equation}\label{5-03-1}
    \alpha^2=5\left(\frac{kT}{\hbar\omega_{ce}}\right)^2.
    \end{equation}
The R.H.S. of the equation describes transitions $\text{E}\leftrightarrow\text{O}$ due to Compton scattering. This equation is solved in the upper layer of the E-mode photosphere, $z\ll\Delta l_{\text E}$, which is transparent for E-photons. Therefore the E-mode radiation density, $\mathcal{E}_\text{E}$, is independent of $z$.
The  equation assumes an isotropic angular distribution of radiation, which is the case for O-photons because the O-mode optical depth is large. The angular distribution of E-photons in the considered layer is not isotropic, since they fall from the fireball in a certain range of angles. However, in addition to incident E-photons, there are also reflected ones, so the difference from the isotropic case is not too large.

When solving  equation~(\ref{00-00}), it is convenient to use the Thomson optical depth $\text{d}\tau=-\sigma_T N_e\text{d}z$. Then the solution with boundary conditions $\mathcal{E}_\text{O}\approx 0$ at $\tau=0$ and $\mathcal{E}_\text{O}\rightarrow\mathcal{E}_\text{E}$ at $\tau\rightarrow\infty$ is written in the form
\begin{equation}\label{00-01}
\mathcal{E}_\text{O}(\tau)= \mathcal{E}_\text{E}\left[1-\exp\left(-\frac{\alpha\tau}{\sqrt{\xi}}\right)\right].
\end{equation}
The gradient of the radiation pressure is $f_\text{rad}=-(1/3)\text{d}\mathcal{E}_\text{O}/\text{d}z$. Thus, the ratio of the radiation to the gravity force is found as
\begin{equation}\label{00-02}
\frac{f_\text{rad}}{f_{\text{gr}}}=\frac{Z\sigma_T}{3Am_pg_*}\sqrt{\frac{5}{\xi}}\frac{kT}{\hbar\omega_{ce}}\mathcal{E}_\text{E}.
\end{equation}
The energy density of the E-mode inside the considered surface layer is determined by the incident radiation flux as $\mathcal{E}_\text{E}\sim F/c$. 
Then the maximal flux of E-mode, above which a static atmosphere is impossible, is estimated as
\begin{equation}\label{00-03}
F_\text{crit}\approx 3\left(\frac{\hbar\omega_{ce}}{kT}\right)F_\text{edd},
\end{equation}
where $F_\text{edd}=g_*c/\kappa_T$ is the classical Eddington flux, and $\kappa_T=N_e\sigma_T/\rho_b$ is the Thomson scattering opacity. Note that the radiation force due to E-mode photons is zero in this case because the total E-mode flux is zero: nearly all the incident radiation is reflected back from the depth $\tau_0\gg\tau_1$. 

\citet{Miller1995}  performed numerical Monte-Carlo simulations of radiation flux  propagated upward
through a magnetar atmosphere and found that the critical luminosity over an area $4\pi R_*^2$ is $L_\text{crit}\approx 5\left(\omega_{ce}/\omega\right)L_\text{edd}$, where $\omega$ is the radiation frequency.  This result was also confirmed by \citet{vanPutten2013}. If we put $\hbar\omega\sim3kT$, rewrite their formula in the terms of the flux, we obtain $F_\text{crit}\approx 1.7(\hbar\omega_{ce}/kT)F_\text{edd}$,
which agrees well with our estimate.
It should be noted that these authors considered the propagation of radiation  upward from the inside atmosphere. In our case, the radiation obliquely falls from above. The similarity of the results is explained by the fact that 
the super-Eddington flux of O-mode radiation in both cases is formed in an upper layer, which is transparent for E-mode radiation. Therefore in both cases, the energy density of E-mode radiation, which is the source of O-mode photons is $\mathcal{E}_\text{E}\sim F/c$. Therefore the flux of O-mode radiation is the same in both cases. A smaller coefficient in their estimate (1.7 instead of our 3) is because in their configuration, the upward E-mode flux contributes to the radiation force whereas in our case, the total E-mode flux is zero because the radiation falls onto the surface and reflected back.


\subsection{The role of photon splitting}

In the previous section, we assumed that the surface is illuminated only by radiation in the E-mode. The fireball does emit only in this mode however, E-photons could split into a pair of O-photons on the way to the surface \citep{Adler1971}. These O-photons propagate in the same direction as the E-photon, from which they originated. 
Approximately one-half of the energy in the E-mode might be transferred to the O-mode via this process. If so, the densities of E- and O-mode radiation remain equal both inside and outside the photosphere so that no radiation pressure gradient is formed. Formally, $\mathcal{E}_\text{O}\approx\mathcal{E}_\text{E}$ both at $\tau\rightarrow\infty$ and at $\tau=0$ so the solution to equation (\ref{00-00}) is $\mathcal{E}_\text{O}\approx\text{const}$.
Thus, material ablation will not occur. 


\begin{figure}
	\includegraphics[width=\columnwidth]{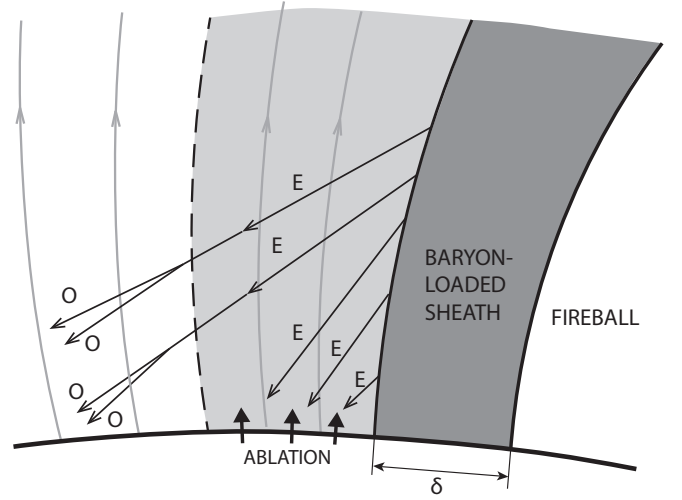}
    \caption{Illumination of the magnetar's surface by radiation from the trapped fireball
    }
    \label{abl}
\end{figure}

The splitting rate for E-photons with the energy $\hbar\omega\ll m_ec^2$ propagating at the angle $\theta$ to the magnetic field $B\gg B_\text{QED}\sin\theta$ is \citep{Adler1971,ThompsonDuncan1995}
\begin{equation}\label{spl}
\Gamma_\text{sp}(\text{E}\rightarrow \text{O}+\text{O})=\frac{\alpha_\text{f}^3}{2160\pi^2}\sin^6\theta\left(\frac{\hbar\omega}{m_ec^2}\right)^5\frac{m_e c^2}{\hbar}.
\end{equation}
Note that this quantity strongly depends on the photon energy and the propagation angle.
After passing the distance $\Delta l$, the E-photon does not split if  $\Gamma_\text{sp}\Delta l/c<1$, which is recast as the condition for the photon energy:
\begin{equation}\label{split1}
\hbar\omega<94\left(\frac{\Delta l}{1\,\text{km}}\right)^{-1/5}\!\!\!\left(\sin\theta\right)^{-6/5} \,\,\,\, \text{keV}.
\end{equation}
 For example, if $\theta\le 45^\circ$, 
 and $\Delta l<1$ km, we get $\hbar\omega<140$ keV. 
It was shown by \citet{ThompsonDuncan1995} that 
the temperature at the outer boundary of the pair fireball is  $T_\text{pair}\approx 30$ keV (more exactly, $T_\text{pair}=27$ keV at $B=10B_\text{QED}$). 
The pair fireball is surrounded by a baryonic sheath therefore the photospheric temperature of the "dressed"\, fireball is even lower. In the thermal emission with such a temperature 
, only a small fraction of the energy is transferred by photons in the tail $\hbar\omega>140$ keV so that splitting does not prevent the evaporation from the region  of the size $l\sim 1-2$ km around the bottom of the fireball (see Fig.~\ref{abl}). In order to reach a larger distance from the fireball, the ray must be emitted close to the normal to the magnetic field, $\theta\sim 90^\circ$. Then photons in the vicinity of the thermal peak, $\hbar\omega\sim (3-4)kT$ could split, so that the energy densities in both modes become comparable, which prevents ablation.

We conclude that the baryonic matter could evaporate only relatively close to the photosphere of the fireball.

\subsection{Formation of a baryon-loaded sheath}

We have shown in the previous subsection that 
the baryonic matter is evaporated from the surface of the neutron star within the distance of 1-2 km from the fireball. The fireball is confined by closed magnetic field lines so that the field lines originating at the evaporating region are typically closed going around the fireball. Therefore the evaporated matter does not escape but accumulates around the fireball forming the baryon-loaded sheath. The matter in the sheath remains suspended in the magnetosphere due to radiation pressure. We will show at the end of this subsection that the sheath of the thickness of 1-2 km is opaque to the E-mode radiation, therefore, a new E-mode photosphere is formed with a lower temperature. If the radiation flux from the new photosphere exceeds the critical value (\ref{00-03}), the baryonic matter is evaporated from the next portion of the neutron star surface. The process continues until the flux from the photosphere drops below the critical value.

Let us estimate the structure of the baryon-loaded sheath. The vertical radiation force in the sheath balances the gravity force therefore the radiation flux along the magnetic field is equal to the magnetically modified flux (\ref{edd2}), $F_\parallel=F_\text{edd}^B$. 
The flux across the sheath is estimated as $F_\perp\sim (l_\parallel/\delta)F_\parallel$, where $\delta$ is the thickness of the sheath and $l_\parallel$ is a characteristic scale on which the radiation energy density changes along the magnetic field. Taking into account that $\delta\le l_{\parallel}$, one concludes that $F_\perp\geq F_\text{edd}^B$.
The magnetically modified Eddington flux is given by equations (\ref{edd}) and (\ref{edd2}); the corresponding temperature is presented by equation (\ref{edd1}). Taking into account equation (\ref{00-03}),
 we obtain, for $Y_e\approx 1$ and $g_*=2\times 10^{14}$ cm/s$^2$,
\begin{equation}\label{crit}
\frac{F_\text{edd}^B}{F_\text{crit}}\approx 2\left(\frac{B(R_*)}{10B_\text{QED}}\right)^{2/3}. 
\end{equation}
This confirms the conjecture of \citet{ThompsonDuncan1995} that the ablation of baryonic matter from the surface of the magnetar occurs when the radiation flux from the fireball is of the order or exceeds the magnetically modified Eddington flux.  


The radiation temperature corresponding to $F^B_{\text{edd}}$ is given by equation (\ref{edd1}). 
For magnetic fields of the order of $10^{14}-10^{15}$ G, the surface temperature of the pair fireball is larger, $T_\text{pair}\sim 30$ keV \citep{ThompsonDuncan1995}. Therefore the width of the baryonic sheath is adjusted such that the temperature falls from $T_{\rm pair}$ to that of equation (\ref{edd1}). 
The radiation flux is constant across the sheath and could be estimated, on the one hand,  as $F_\perp\sim c\mathcal{E}_\text{E,in}/\tau_\perp$, and on the other hand, as 
$F_\perp\sim c\mathcal{E}_\text{E,out
}$, where $\mathcal{E}_\text{E,in}$ and $\mathcal{E}_\text{E,out}$ are the energy density of the E-mode at the internal boundary of the sheath and on its surface, respectively. Therefore the optical depth of the baryon-loaded sheath, with respect to E-mode radiation is $\tau_\perp\sim \mathcal{E}_\text{E,in}/\mathcal{E}_\text{E,out}\sim(T_\text{pair}/T)^4$. In the stationary case, when $F_\perp\sim F_\text{edd}^B$ the temperature of the E-mode photosphere is $kT\sim 16$ keV (see equation~(\ref{edd1})), therefore $\tau_\perp\sim 10$.

Let us estimate the geometrical thickness, $\delta$, of such a sheath. Considering the equilibrium, when the gravity force is equal to the radiation force, the plasma density can be estimated as
\begin{equation}\label{dens0}
\rho_b\approx-\frac{1}{g}\frac{\text{d}P_\text{rad}}{\text{d}R}\sim-\frac{16\sigma_{\text{SB}}T^3}{3cg}\frac{\text{d}T}{\text{d}R}.
\end{equation}
The temperature across the sheath varies as  $T^4\approx T_\text{pair}^4/\tau_\perp$, and the vertical scale is of the order of the star radius. 
Therefore, we obtain the following equation for the optical depth
\begin{equation}\label{dens}
\text{d}\tau_\perp=\sigma_R N_e\text{d}\delta\sim\frac{80\pi^2}{3} \frac{\sigma_\text{SB} T_\text{pair}^4}{F_\text{edd}}\left(\frac{kT_\text{pair}}{\hbar\omega_{ce}}\right)^2\tau_\perp^{-3/2}\frac{\text{d}\delta}{R_*}.
\end{equation}
The solution of this equation can be written as
\begin{equation}\label{dens1}
\frac{\delta}{R_*}\approx 0.5 \left(\frac{B(R_*)}{10B_\text{QED}}\right)^2\left(\frac{kT_\text{pair}}{27\,\,\text{keV}}\right)^{-6}\left(\frac{\tau_\perp}{10}\right)^{5/2}.
\end{equation}
For $\tau_\perp\sim 10$, we obtain that $\delta$ is of the order of $R_*/2$, which agrees with the conclusion of \citet{ThompsonDuncan1995} that the thickness of the baryon-loaded sheath is comparable with the size of the star. 

At the beginning of this subsection, we described the step-by-step formation of the baryonic sheath. We assumed that the initial sheath of the thickness $\delta\sim 1-2$ km has the optical depth exceeding unity such that the new photosphere is formed at the outer boundary of the sheath. We now see from equation (\ref{dens1}) that this assumption is fulfilled, $\tau_\perp>1$ for $\delta/R_*\sim 0.1-0.2$.


\section{Outflow of baryonic plasma}

It was shown in the previous section that the baryonic plasma is evaporated from a large region of the magnetar's surface around the fireball.  
It is possible that this region includes the polar cap from which the magnetic field lines go to infinity. In this case, the magnetar wind is loaded by baryons during the flare. In this section, we estimate the mass flow in such a wind. 

The plasma outflow is described by the equations of radiation hydrodynamics. 
There are two sound points in this case. The first one corresponds to the gas speed of sound, $c_s^2\sim kT/m_p$, 
and the second one to the radiative speed of sound, $c_{\text{rs}}^2=\text{d}\mathcal{E}_\text{O}/\text{d}\rho_b$. 
The gas speed of sound is very small as compared with the free fall velocity, therefore, the flow passes the gas sound point  
near the surface of the magnetar, where the density is large and the radiation energy densities in both modes are nearly equal. 
Therefore, we can consider this sound point as the lower boundary in our problem. 
The passage through the second sound barrier is possible only in an expanding flow, so we have to take into account the curvature of the magnetic field lines. 

Note that the exact solution depends on the geometry of the magnetic field lines, which is hardly ever dipolar in an active magnetar. Moreover, the free-fall velocity at the surface of the neutron star is mildly relativistic, therefore, strictly speaking, one has to solve equations of relativistic radiation hydrodynamics. Having in mind obtaining just order of magnitude estimates and rough scalings, we neglect relativistic effects and use the dipole geometry.

\subsection{General equations}
We consider  a stationary plasma outflow  along the magnetic field lines. Indeed, magnetic energy density is much higher than kinetic energy density of plasma, and they are comparable only at a sufficiently large distance from the magnetar (see Section 4). In this case, the plasma can only flow along magnetic field lines, therefore, the Lorentz force can be ignored. The continuity equation and the equation of motion have the form 
\begin{gather}\label{3-02}
\nabla\cdot(\rho_b\mathbf{v}_b)=0;\\
\label{3-01}
\rho_b(\mathbf{v}_b\cdot\nabla)\mathbf{v}_b=-\nabla p_\text{pl}+\rho_b\mathbf{g}+\rho_b\mathbf{f}_\text{rad}
\end{gather}
where $\mathbf{v}_b$ is the plasma velocity,  $\rho_b\mathbf{f}_\text{rad}$ -- the radiation force, $\mathbf{g}$ -- the gravity acceleration. 


As it was demonstrated in Section 2.2, the outflow is produced by the super-Eddington flux of the O-mode radiation. Since the optical depth for this radiation is large, we can use the diffusion approximation. Then the radiation force is written as
\begin{equation}\label{3-01-1}
\mathbf{f}_{\text{rad}}=-\frac{1}{3\rho_b}\nabla\mathcal{E}_\text{O}
\end{equation}
where $\mathcal{E}_\text{O}$ is the energy density of O-photons. 

The flow in a narrow open field line tube is transparent to E-photons. Moreover, close to the star's surface, where most of acceleration takes place, the flux of the E-mode radiation vanishes because the radiation falls from above and is reflected back. At the altitudes comparable and larger than the stellar radius, the overall E-mode flux is directed outwards. Being comparable with the magnetically modified Eddington flux it could accelerate the flow. We neglect this effect because the outflow velocity reaches the free-fall velocity at the altitude comparable with the stellar radius anyway. Adding the force comparable with the gravity force could not affect the result qualitatively. 

The  transfer equation for the O-mode radiation is written as the energy density equation (see Appendix B)
\begin{equation}\label{5-03}
\begin{split}
(\mathbf{v}_b\cdot\nabla) \mathcal{E}_\text{O}-\nabla\cdot\left(D_\text{O}\nabla \mathcal{E}_\text{O}\right) & +\frac{4}{3}\mathcal{E}_\text{O}(\nabla\cdot \mathbf{v}_b)=\\
& =c\sigma_TN_e\alpha^2\frac{\left(Q^{3/2}-\mathcal{E}^{3/2}_\text{O}\right)}{Q^{1/2}};
\end{split}
\end{equation}
where $\alpha$ is defined by the equation (\ref{5-03-1}).
In this equation, the first term in the L.H.S. describes advection, the second -- diffusion, and the third -- the adiabatic energy variation. The R.H.S describes the energy exchange between the E- and O-modes.
The quantity $Q$ is the rate of the O-photon energy production due to the conversion from the E-mode to the O-mode (if E-mode photons have isotropic angular distribution, then 
$Q=\mathcal{E}_\text{E}$ ). The energy exchange between the modes is significant only close to the stellar surface, where the plasma density is large. Since the plasma is transparent for E-photons, their density is constant in this region, therefore we can assume that $Q\approx\text{const}$. 



Let us introduce dimensionless quantities: 
\begin{equation}\label{6-01}
\begin{split}
&Q=\frac{m_p g}{\alpha^2\pi^2 Y_e\sigma_T}q, \quad \mathcal{E}_\text{O}=\frac{m_p g}{\alpha^2\pi^2 Y_e\sigma_T}\mathcal{E}, \\
&\mathbf{v}_b=v_\text{esc}\mathbf{u}, \quad  \rho_b = \frac{m_p g}{\alpha^2\pi^2 v_\text{esc}^2Y_e\sigma_T}\rho;
\end{split}
\end{equation}
where $v_\text{esc}=\sqrt{2g_*R_*}$ is the free-fall velocity. It is easy to see that $q\sim1$ corresponds to the magnetically modified Eddington flux, $Q\sim F_\text{edd}^B\sim 10^{29}$ erg/cm$^2$s (see equation~(\ref{edd})), and corresponding luminosity over an area $4\pi R_*^2$ is  $L_\text{edd}^B\sim 10^{42}$ erg/s.
The dimensionless equations are written as
\begin{gather}\label{6-02}
\nabla\cdot(\rho\mathbf{u})=0,\\
\label{6-03}
\rho(\mathbf{u}\cdot\!\nabla)\mathbf{u}=-\nabla\left(\frac{p_\text{pl}}{m_pv_\text{esc}^2}\right) +\rho\frac{\mathbf{g}R_*}{v_\text{esc}^2}-\frac{1}{3}(\mathbf{n}\cdot \nabla)\mathcal{E},
\\ \label{6-04}
(\mathbf{u}\cdot\!\nabla)\mathcal{E}-\!\frac{2\pi^2\alpha^2\xi}{\beta_\text{esc}}\nabla\left(\frac{1}{\rho}\nabla\mathcal{E}\right)\!+\!\frac{4}{3}\mathcal{E}(\nabla\cdot\mathbf{u})=\frac{\rho(q^{3/2}\!-\mathcal{E}^{3/2})}{2\pi^2\beta_\text{esc}q^{1/2}},
\end{gather}
where the operator $\nabla$ is defined with respect to dimensionless coordinates (which we normalize to the magnetar radius $R_*$) and $\beta_\text{esc}=v_\text{esc}/c$. It should be noted that the small parameter $\alpha^2$ enters only into the term characterizing the diffusion of photons.

\subsection{Scaling relation for the plasma mass flow}

In this subsection we roughly estimate the dependence of the mass flow, $j$, on the energy flux of the incident radiation, $q$. In the next subsection, we present the numerical solution, which confirms the obtained scaling.
Let us analyse equations at not too high altitudes, $x< 1$, but above the gas sonic point, so that we can neglect the expansion of the flow and the gas pressure. We also neglect the radiation diffusion, which will be justified a posteriori. 
In this case,  equation (\ref{6-02}) implies $\rho u=j$. Eliminating $\rho$ from equations (\ref{6-03}) and (\ref{6-04}) yields
\begin{gather}
j\frac{du}{dx}=-\frac{j}{2u}-\frac{1}{3}\frac{d\mathcal{E}}{dx},
\\
u\frac{d\mathcal{E}}{dx}+\frac{4}{3}\mathcal{E}\frac{du}{dx}=
\frac{j(q^{3/2}-\mathcal{E}^{3/2})}{2\pi^2
\beta_\text{esc}q^{1/2}u}
\end{gather}
In order to get an approximate analytical solution,
we substitute the expression $(q^{3/2}-\mathcal{E}^{3/2})/q^{1/2}$ by a simpler expression $(q-\mathcal{E})$, which has the same asymptotics both at $\mathcal{E}\approx q$ and at $\mathcal{E}\ll q$. 
Now the system of equations can be rewritten in the following form
\begin{equation}\label{7-07}
\frac{\text{d}u}{\text{d}x}=\frac{1}{3}\left(\frac{q-\mathcal{E}}{2\pi^2\beta_\text{esc}u}+\frac{3}{2}\right) \left(\frac{4\mathcal{E}}{9j}-u\right)^{-1}
\end{equation}
\begin{equation}\label{7-08}
\frac{\text{d}\mathcal{E}}{\text{d}x}=-\left(\frac{j(q-\mathcal{E})}{2\pi^2\beta_\text{esc}u}+\frac{2\mathcal{E}}{3u}\right) \left(\frac{4\mathcal{E}}{9j}-u\right)^{-1}
\end{equation}
This set of equations has the singular point, $x_\text{rs}$, such that $u(x_\text{rs})=4\mathcal{E}(x_\text{rs})/9j$. At this point, the plasma velocity is equal to the local radiative sound speed, $c_\text{rs}=\sqrt{\text{d}\mathcal{E}_\text{O}/\text{d}\rho_b}=\sqrt{\Gamma_a\mathcal{E}_\text{O}/3\rho_b}$, where $\Gamma_a\approx 4/3$ is the adiabatic index in the radiation-pressure dominated limit (\citealt{Chandrasekhar1967}). This speed differs from the relativistic speed of sound $c/\sqrt{3}$, because the plasma is not relativistic, i.e. $kT\ll m_e c^2$. As can be seen, the numerator in the above equations is always strictly positive or negative and never vanishes in the planar geometry. 
In order to obtain a solution passing the singular point, one must take into account that the outflow expands; then the numerator explicitly depends on $x$.

In order to find a rough estimate, we will use the fact that the sonic point occurs at the altitude comparable with the stellar radius, $x_\text{rs}\sim 1$, and the velocity of the flow at this point is comparable with the escape velocity, $u(x_\text{rs})\sim 1$. On the other hand, most of the acceleration occurs at $x\ll 1$. Therefore we find the dependence of the mass outflow on the incident radiation flux by extrapolating the solution to equations (\ref{7-07}) and (\ref{7-08}) to $u\sim 1$.

Dividing these equations one by the other, we obtain the equation for the function $u(\mathcal{E})$
\begin{equation}\label{10-03}
\frac{\text{d}u}{\text{d}\mathcal{E}}=-\frac{q-\mathcal{E}+3\pi^2\beta_\text{esc}u}{3j(q-\mathcal{E})+4\pi^2\beta_\text{esc}\mathcal{E}}
\end{equation}
The solution to this linear differential equation has the form
\begin{equation}\label{10-03-1}
u(\mathcal{E})=C(\mathcal{E})\left[3j(q-\mathcal{E})+4\beta\mathcal{E}\right]^{\frac{3\beta}{3j-4\beta}},
\end{equation}
where 
\begin{equation}\label{10-03-2}
C(\mathcal{E})=C+\frac{(7q-3\mathcal{E})}{3(3j-7\beta)}[3j(q-\mathcal{E})+4\beta\mathcal{E}]^{-\frac{3\beta}{3j-4\beta}}.
\end{equation}
Here we introduced the notation $\beta=\pi^2\beta_\text{esc}$.
The constant $C$ is found from the condition $u_0= 0$ at $\mathcal{E}_0= q$. Then the solution is written as
\begin{equation}\label{10-03-3}
u(\mathcal{E})= \frac{7q-3\mathcal{E}}{3(3j-7\beta)}-\frac{4q}{3(3j-7\beta)}\left[\frac{3j}{4\beta}\!\left(1-\frac{\mathcal{E}}{q}\right)+\frac{\mathcal{E}}{q}\right]^{\frac{3\beta}{3j-4\beta}}.
\end{equation}
Extrapolating equation (\ref{10-03-3}) to the escape velocity, $u\sim 1$, and taking into account that in this region,  the flow has already expanded significantly, so that $\mathcal{E}\ll q $, 
we obtain the dependence $q(j)$: 
\begin{equation}\label{10-05}
q(j)= 3(3j-7\beta)\left[7-4\left(\frac{3j}{4\beta}\right)^{\frac{3\beta}{3j-4\beta}}\right]^{-1}
\end{equation}
In the limit $j\ll\beta\sim 2\pi^2/3$, this relation is simplified to the form
\begin{equation}\label{10-05-1}
j= 0.1q^{4/3}
\end{equation}
One sees that for $q\le 10$, the condition $j\ll\beta$ is satisfied. 
The scaling (\ref{10-05-1}) will be  confirmed by the numerical solution.

Now let us show that the diffusion radiation flux could be neglected as compared with the advection flux.  The ratio of fluxes, i.e., the ratio of the second to the third terms in the equation~(\ref{6-04}) is written as:
\begin{equation}\label{flux_ratio}
    \frac{F_\text{diff}}{F_\text{adv}}=-\frac{3\pi^2\alpha^2\xi}{2\beta_\text{esc}j}\frac 1{\mathcal{E}}\frac{d\mathcal{E}}{dx}.
\end{equation}
One sees that the ratio decreases with the increasing altitude, $x$, therefore it is sufficient if we show that this ratio is small already at small $x$.

Close to the surface, $u\ll 1$  and $q-\mathcal{E}\ll q$ therefore equation (\ref{10-03-3}) is reduced to \begin{equation}\label{10-03-4}
u\approx \frac{(q-\mathcal{E})^2}{8\beta q}.
\end{equation}
Substituting this relation to equations (\ref{7-07}) and (\ref{7-08}) and expanding in small parameters yields a simple solution
\begin{equation}\label{10-03-5}
    u=\frac{(36j x)^{2/3}}{8(\beta q)^{1/3}}; \quad  \mathcal{E}=q-(36j\beta q x)^{1/3}.
\end{equation}
This solution is valid when the velocity of the flow velocity exceeds the gas speed of sound, $u^2\gg kT/(m_pv^2_\text{esc})$, i.e. at
\begin{equation}\label{10-03-6}
    x\gg x_0=\frac{2^{5/2}q^{1/2}\pi}{9j\beta_\text{esc}}\left(\frac{kT}{m_p c^2}\right)^{3/4}.
\end{equation}
On the other hand, the assumed condition $q-\mathcal{E}\ll q$ implies $x\ll q^2/(36j\beta)\sim 0.1$ (note that $j\sim 0.1$ according to the equation~(\ref{10-05-1})). 
Substituting the obtained solution into the equation~(\ref{flux_ratio}) yields
\begin{equation}\label{10-03-7}
\begin{split}
    \frac{F_\text{diff}}{F_\text{adv}}&= \frac{3^{2/3}\pi^{8/3}\xi\alpha^2}{2^{1/3}(\beta_\text{esc}jqx)^{2/3}}= \\
    &=0.4\,\frac{\xi}q\left(\frac{kT}{20\,\text{keV}}\right)^{3/2}\left(\frac{10B_\text{QED}}{B}\right)^{2}\left(\frac{x_0}x\right)^{2/3}.
\end{split}
\end{equation}
This estimate justifies neglecting the radiation diffusion well above the gas sonic point.

\subsection{Numerical solution}

\begin{figure}
	\includegraphics[width=\columnwidth]{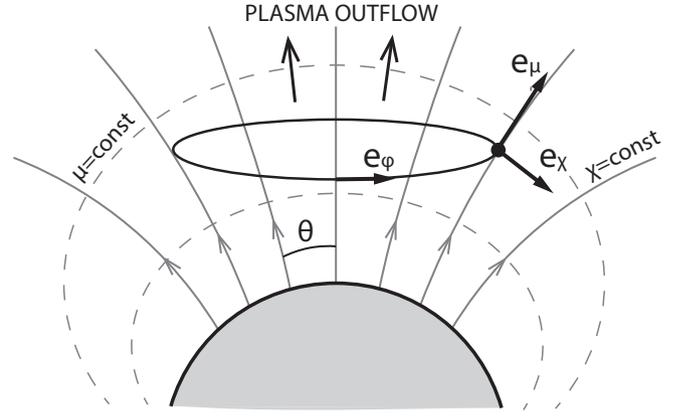}
    \caption{Dipolar coordinates}
    \label{dipolar}
\end{figure}

It was shown in the previous subsection that in the planar case the outflow cannot pass the sonic point; one has to take into account the expansion of the flow. The geometry of the outflow depends on the structure of the magnetic field, which could be quite complicated near the surface of an active magnetar.
To be specific, let us consider the outflow near the magnetic axis of the dipole field. We define the right-handed dipole coordinate system $(\mu,\chi,\phi)$, which is resented in spherical coordinates, $(r,\theta,\phi)$, by
\begin{equation}\label{8-01}
\mu=-\frac{\cos\theta}{r^2}, \quad \chi=\frac{\sin^2\theta}{r}, \quad \phi=\phi ,
\end{equation}
where $r=R/R_*$ is the dimensionless radial coordinate. The differential operators are written in this coordinate system in Appendix C.
The plasma flows along the magnetic field lines, $\mathbf{u}=u\mathbf{e}_\mu$, where $\mathbf{e}_\mu$ is the unit vector along the coordinate line $\chi=\text{const}$, $\phi=\text{const}$ (see Fig.~\ref{dipolar}). We consider open field lines tube near polar caps when $\theta\ll 1$. 

Projecting the equation~(\ref{6-03}) onto the $\mu$ axis at $\theta\rightarrow 0$,  we obtain 
\begin{equation}\label{8-02}
\rho u\frac{\text{d} u}{\text{d} r}=- \frac{\rho}{2r^2}-\frac{1}{3}\frac{\text{d} \mathcal{E}}{\text{d} r}.
\end{equation}
The continuity equation in dipole coordinates at $\theta\rightarrow 0$ can be written as 
\begin{equation}\label{8-04}
r^3\rho u=j=\text{const}.
\end{equation}
We demonstrated in the previous subsection, that well above the gas sound point, we can neglect diffusion 
with respect to advection in the energy equation~(\ref{6-04}). Then we obtain in the dipole coordinate system
\begin{equation}\label{8-08}
u\frac{\text{d}\mathcal{E}}{\text{d}r}+\frac{4}{3}\mathcal{E}\frac{1}{r^3}\frac{\text{d}}{\text{d}r}\left(r^3 u\right)=\frac{\rho \left(q^{3/2}-\mathcal{E}^{3/2}\right)}{2\pi^2\beta_\text{esc}q^{1/2}}.
\end{equation}

Eliminating the density from the resulting set of equations, we can rewrite it in a more convenient form by explicitly expressing the first derivatives of $u$ and $\mathcal{E}$ 
\begin{gather}\label{8-09}
\frac{\text{d}u}{\text{d}r}=\frac{r^3}{3j} \left(\frac{j(q^{3/2}-\mathcal{E}^{3/2})}{2\pi^2\beta_\text{esc}ur^3q^{1/2}}-4\mathcal{E}\frac{u}{r}+\frac{3j}{2r^5}\right) \left(\frac{4\mathcal{E}r^3}{9j}-u\right)^{-1},
\\ \label{8-10}
\frac{\text{d}\mathcal{E}}{\text{d}r}=-\left(\frac{j(q^{3/2}-\mathcal{E}^{3/2})}{2\pi^2\beta_\text{esc}ur^3q^{1/2}}-4\mathcal{E}\frac{u}{r}+\frac{2\mathcal{E}}{3ur^2}\right) \left(\frac{4\mathcal{E}r^3}{9j}-u\right)^{-1}.
\end{gather}
In these equations, the denominator vanishes at the singular point $r_\text{rs}$, just as in the planar case. However, a smooth solution exists because now the numerator explicitly depends on $r$, so that it can be set to zero at this point. 
Therefore, the parameters at the singular point satisfy the relations \begin{equation}\label{8-11}
u_\text{rs}=\frac{4r_\text{rs}^3}{9j}\mathcal{E}_\text{rs}, \,\,\,\,\, \frac{q^{3/2}-\mathcal{E}^{3/2}_\text{rs}}{6\pi^2\beta_\text{esc}u_\text{rs}q^{1/2}}-\frac{4}{3j}\mathcal{E}_\text{rs} u_\text{rs} r_\text{rs}^2+\frac{1}{2r_\text{rs}^2}=0
\end{equation}
The first equation corresponds to vanishing the denominators, and the second one corresponds to vanishing the numerators (in both equations (\ref{8-09}) and~(\ref{8-10})). 
Equations~(\ref{8-11}) must be solved together with differential equations (\ref{8-09}) and (\ref{8-10}). 

We solve these equations numerically as follows.
By choosing two parameters, $r_\text{rs}$ and $j$,
we find $u_\text{rs}$ and $\mathcal{E}_\text{rs}$ from equations (\ref{8-11}). Then equations (\ref{8-09}) and (\ref{8-10}) can be solved starting from the vicinity of the singular point and integrating numerically inwards and outwards. The parameters $r_\text{rs}$ and $j$ are found by the shooting method such that $u\to 0$ and $\mathcal{E}\to q$ at $r\to 1$. 

The integration starts at the points $r_\text{rs}\pm \text{d}r$, where we choose $\text{d}r=10^{-6}$. Inasmuch as the solution smoothly passes the singular point, one can write
\begin{equation}\label{10-01}
u(r_\text{rs}\pm\text{d}r)\approx u_\text{rs}\pm\left(\frac{\text{d}u}{\text{d}r}\right)_\text{rs}\!\!\text{d}r, \,\,\, \mathcal{E}(r_\text{rs}\pm\text{d}r)\approx \mathcal{E}_\text{rs}\pm\left(\frac{\text{d}\mathcal{E}}{\text{d}r}\right)_\text{rs}\!\!\text{d}r.
\end{equation}
The derivatives of the unknown functions at the singular point are found  by taking the limit $r\rightarrow r_\text{rs}$ in the equations~(\ref{8-09}) and~(\ref{8-10}) and using  L'Hopital's rule since we get the uncertainty $0/0$ in the R.H.S. Then we obtain a system of equations for the derivatives $(\text{d}u/\text{d}r)_\text{rs}$ and $(\text{d}\mathcal{E}/\text{d}r)_\text{rs}$ . This system has several solutions, but we choose the one where $(\text{d}u/\text{d}r)_\text{rs}>0$ and $(\text{d}\mathcal{E}/\text{d}r)_\text{rs}<0$. Finding the derivatives, we find the initial values, $u(r_\text{rs}\pm \text{d}r)$ and $\mathcal{E}(r_\text{rs}\pm \text{d}r)$ for the integration.



\begin{figure}
	\includegraphics[width=\columnwidth]{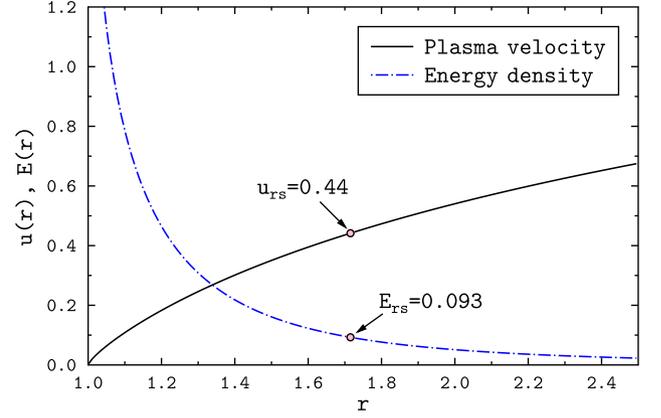}
    \caption{The plasma velocity, $u(r)$, and the O-photon energy density, $\mathcal{E}(r)$, for $q=3$. }
    \label{plot1}
\end{figure}

\begin{figure}
	\includegraphics[width=\columnwidth]{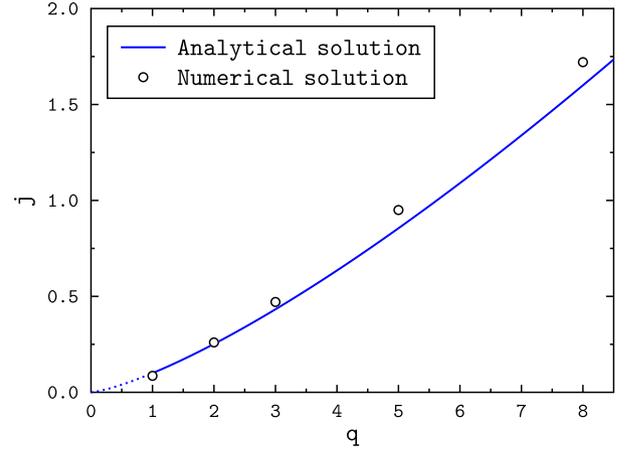}
    \caption{The dependence of the mass flow on the parameter $q$ evaluated from the analytical estimate  (\ref{10-05-1})  (solid curve) and according to the numerical solution (circles)}
    \label{fig:list2}
\end{figure}

The results of simulations for different values of the radiation parameter $q\geq1$ are presented in Table 1. In Fig. \ref{plot1}, we present the solution for $q=3$. The dependence of the mass flux on 
$q$ is presented in Fig. \ref{fig:list2}. One sees that the analytical estimate  (\ref{10-05-1}) describes the numerical results quite well.

Note that the neglect of the diffusion as compared with the advection was justified in the previous subsection only for plane flows. In an expanding flow, the  optical depth decreases so that eventually the diffusion may become significant. However, the flow expands only a few times at altitudes of the order of the stellar radius, where the radiation sonic point is passed. Therefore the mass outflow is established by the solution at moderate altitudes, when the diffusion is still neglected. Numerical estimate of the diffusion flux confirms this conjecture.

The flow velocity at the singular point (which is equal to the radiation speed of sound), $u_\text{rs}$, is less than $v_\text{esc}/c=\sqrt{2g_*R_*}/c\approx 0.67$, but close to it. However, an expanded supersonic flow accelerates during expansion, 
so that the plasma velocity can easily reach $v_\text{esc}$ if it moves along the opened magnetic field lines. Then the plasma escapes from the magnetar's magnetosphere forming a baryonic wind. 




\begin{table}
\caption{Results of the numerical simulations}
\label{tab1}
\begin{tabular}{lcccc}
\hline
$q$ & $j$ & $r_\text{rs}$ & $u_\text{rs}$ & $\mathcal{E}_\text{rs}$\\
\hline
$1$ & $8.6\times 10^{-2}$ & $2.16$ & $0.36$ & $7\times 10^{-3}$\\
$2$ & $0.26$ & $1.85$ & $0.41$ & $3.8\times 10^{-2}$\\
$3$ & $0.47$ & $1.72$ & $0.44$ & $9.3\times 10^{-2}$\\
$5$ & $0.95$ & $1.59$ & $0.49$ & $0.26$\\
$8$ & $1.72$ & $1.51$ & $0.54$ & $0.61$ \\
\hline
\end{tabular}
\end{table}

\section{A baryon dominated wind}

Let us consider the motion of plasma along open magnetic field lines. 
It follows from equations~(\ref{8-11}),
that at the radiation sonic point, $(4/3)\mathcal{E}_\text{rs}r_\text{rs}^3=3ju_\text{rs}$. Then the total energy flux at this point is  presented as
\begin{equation}F_\text{tot}= \frac 12ju^2+\frac 43\mathcal{E}ur^3=\frac 72ju^2
\end{equation}
so that the ratio of the total energy flux to the plasma kinetic energy flux, $F_\text{kin}=(1/2)ju^2$, is equal to 7 independently of the parameter $q$.  In an expanded flow, the total energy is eventually converted to the kinetic energy, so that $F_\text{tot}=(\Gamma_\text{max}-1)j(c/v_\text{esc})^2$,
where $\Gamma_\text{max}$ is the final Lorentz factor of the flow. This yields $\Gamma_\text{max}\approx 1+(7/2)u_\text{rs}^2(v_\text{esc}/c)^2$. For the parameters of interest, the flow becomes mildly relativistic, $\Gamma_\text{max}\sim 1$. 

In this estimate, we neglected the radiation force due to the E-mode radiation from the fireball, as well as the gravity force. These forces are counteracting and of the same order because the E-mode flux is of the order of the magnetically modified Eddington flux. Therefore these forces could accelerate or decelerate the flow not larger than the free-fall velocity, which does not affect our conclusion that the flow is mildly relativistic. 

The plasma flow does not affect the magnetospheric magnetic field as soon as the plasma kinetic energy density remains smaller than the magnetic energy density. This condition is violated at the Alfvén radius, beyond which the magnetosphere becomes open and the plasma is ejected from the magnetosphere forming a magnetized wind. Assuming the dipole magnetospheric field and substituting $\Gamma=\Gamma_\text{max}\sim 1$ and parameters from equations~(\ref{6-01}) and (\ref{10-05-1}) into the relation
\begin{equation}\label{13-01}
\Gamma \rho_b c^2 \sim \frac{B^2}{8\pi},
\end{equation}
we estimate the Alfvén radius as
\begin{equation}\label{13-03}
\frac{R_A}{R_*}\approx\left(\frac{B_*^2}{8\pi}\frac{\alpha^2\upi^2Y_e\sigma_T}{m_p gc}\frac{v_\text{esc}}{j\Gamma_\text{max}}\right)^{1/3}\!\!\!\!\! \sim 3.4\times 10^3\left(\frac{kT}{20\,\text{keV}}\right)^{2/3}\!\!\!\!\!q^{-4/9}.
\end{equation}
One sees that the Alfvén radius is smaller than the radius of a light cylinder, $R_L/R_*=c/(\Omega R_*)=4.77\times 10^3 P$, for a typical magnetar period $P\sim 5$-$10$ seconds. 

The radius of the open filed line tube at the surface of the star is $a=R_*(R_*/R_A)^{1/2}$. 
Therefore the total mass outflow is estimated, with account of equations~(\ref{6-01}), (\ref{10-05-1}) and (\ref{13-03}), as
\begin{equation}\label{massflow}
    \dot M\!=\!\pi a^2\!\rho_bv_b\!\approx\! 7\!\times\! 10^{15}\!\left(\frac{B(R_*)}{10B_\text{QED}}\right)^2\!\!\left(\frac{kT}{20\,\text{keV}}\right)^{-8/3}\!\!\!\!q^{16/9}\!\!  \quad \text{g/s}
\end{equation}
One sees that the mass flow varies from $10^{16}$ g/s to $10^{18}$ g/s, if $q$ varies from $1$ to $10$. The value of $q$, in turn, depends on how far the open magnetic field lines are from the fireball. The closer the fireball is to open magnetic field lines, the greater the parameter $q$ on this line is, and the greater the mass flow $\dot M$ we obtain. Our results is consistent with the inequality for the mass flow $\dot M<
(\pi/\Delta\Omega_\text{jet})L_\text{edd}(GM/R_*)^{-1}\sim 10^{19}$ g/s, which provides collimation of X-ray radiation jets during giant flares \citep{ThompsonDuncan2001,vanPutten2016}.

The total mass ejected along open magnetic field lines is  estimated as
\begin{equation}\label{mass}
    M\sim 7\times 10^{17}\left(\frac{\tau}{100\,\text{s}}\right)\!\left(\frac{B(R_*)}{10B_\text{QED}}\right)^2\!\!\left(\frac{kT}{20\,\text{keV}}\right)^{-8/3}\!\!\!\!q^{16/9}\!\!  \quad \text{g}.
\end{equation}
where $\tau\sim 100$ s is the duration of a giant flare. One can see that $M$ is much smaller than total mass of mildly relativistic baryonic cloud, $M\gtrsim10^{24.5}$ g,
which was estimated using the radio afterglow for the giant flare from SGR 1806-20 \citep{Gaensler05,Gelfand_etal05,Granot_etal06}. This means that this cloud has been ejected during the hard spike when the luminosity significantly exceeds even the magnetically modified Eddington luminosity and the  magnetosphere becomes in fact open. Here we consider evaporation in the course of intermediate flares and during the fireball stage of giant flares, when most of the magnetosphere remains closed. 

Beyond the Alfvén radius, the magnetic field is wrapped backward because of the plasma inertia. Therefore, the magnetic field in the wind is helical. 
Up to the Alfvén radius, the plasma rigidly rotates with the magnetosphere therefore the rotational velocity at the Alfvén radius is $\sim\Omega R_A$. The radial velocity is of the order of the speed of light because the flow is mildly relativistic.   The frozen-in condition, 
\begin{equation}\label{13-04}
\mathbf{E}+(1/c)\mathbf{v}\times\mathbf{B}=0
\end{equation}
implies that at the  Alfvén surface, 
$E\sim (\Omega R_A/c) B$. The magnetic field is frozen into the plasma, therefore the ratio of the azimuthal to the poloidal components of the magnetic field is of the order of the ratio of the rotational to the radial velocities, $B_{\varphi}/B_p\sim \Omega R_A/c$. 
Now the ratio of the Poynting to the kinetic energy flux is found as
\begin{equation}\label{13-05}
\sigma=\frac{EB_{\varphi}}{4\pi\Gamma\rho_bc^2}
\sim\left(\frac{B_\phi}{B}\right)^2\sim \left(\frac{R_A}{R_L}\right)^2 \sim 10^{-2}.
\end{equation}
Here the equation (\ref{13-01}) was used. One concludes that the baryonic wind is weakly magnetized and mildly relativistic.

The ejected matter takes away the angular momentum $\sim\Omega R_A^2$ per unit mass. Therefore torque applied to the neutron star is estimated as $K\sim R_A^3B_\varphi B_p\sim 8\pi\rho_bcR_A^4\Omega$, where all quantities are estimated at the Alfvén radius. Without the baryonic outflow, the torque is determined by the parameters at the light cylinder radius. Taking into account that the field at the Alfvén radius exceeds the field at the light cylinder in the normal state roughly $(R_L/R_A)^3\sim 10^3$ times, one concludes that the spindown rate at the stage of the baryonic wind exceeds the "normal" spindown rate roughy $100$ times. Specifically, the spindown rate is estimated as
\begin{equation}
    \dot P=\frac{KP^2}{2\pi I}\sim 6\times 10^{-9}\frac{P}{10\, \text{s}}\left(\frac{B(R_*)}{10B_\text{QED}}\right)^2\left(\frac{kT}{20\,\text{keV}}\right)^{-4/3}\!\!\!q^{8/9}.
\end{equation}
Here we used the momentum of inertia of the star, $I=10^{45}$ g$\cdot$cm$^2$. For the duration of a giant flare,
the magnetar period increases by
\begin{equation}
    \frac{\Delta P}{P}\sim 6\times 10^{-8}\left(\frac{\tau}{100\, \text{s}}\right)\left(\frac{B(R_*)}{10B_\text{QED}}\right)^2\left(\frac{kT}{20\,\text{keV}}\right)^{-4/3}\!\!\!q^{8/9}.
\end{equation}
The resulting period increment $\sim6\times 10^{-8}$ is consistent with the upper limit $\Delta P/P<5\times 10^{-6}$ found for the December 27 giant flare, which is associated with SGR 1806-20 source (\citealt{Woods2007}). However, in the August 27 giant flare from SGR 1900+14, a very large positive period increment was observed, $\Delta P/P\sim 10^{-4}$ (\citealt{Woods1999}). Such an increment could not be caused by a plasma outflow during the flare. 
\citet{Thompson2000} have shown that even if the energy of the ejected mass is comparable with the total energy of the flare, the resulting increase of the period is still well below the observed. They attribute the observed period increment to 
the exchange of the angular momentum between the neutron superfluid and the rest of the star.

\section{Conclusions}

In this paper, we considered the evaporation of plasma from the magnetar's surface illuminated by the powerful radiation from the trapped fireball in course of magnetar flares. We confirm the picture envisioned by \citet{ThompsonDuncan1995}: the evaporated plasma forms a baryon-loaded sheath around the fireball. We have shown that the main mechanism of evaporation is the scattering of the E-mode radiation into the O-mode within a dense surface layer of the star. The super-Eddington flux of the O-mode radiation pushes the matter upwards. The E-photon splitting into O-photons on the way to the surface could prevent evaporation because if a large enough O-mode radiation density is produced above the surface, no radiation pressure gradient is formed necessary to uplift the matter. Therefore the matter is evaporated only in the vicinity of the fireball, where the E-photons have not split yet. When a narrow sheath is formed, the new photosphere emits E-mode radiation, which ablates a new portion of the matter forming the next baryonic layer. The process continues until the radiation flux from the fireball drops below the magnetically modified Eddington flux. 

The width of thus formed baryon-loaded sheath is large, about one-half of the stellar radius. If the magnetic pole finds itself within such a sheath, a mildly relativistic, baryonic wind is formed. We estimated parameters of this wind. The wind is weakly magnetized however, it produces the torque roughly two orders of magnitude larger than the "normal" torque. Still, because of small duration of flares, the resulting increase in the magnetar period is very small. 

\section*{Acknowledgements}

This research was supported by the Israel Science Foundation under the grant 2067/19.

\section*{Data Availability}

No new data were generated or analysed in support of this research.


\bibliographystyle{mnras}
\bibliography{References} 




\appendix

\section{Vacuum resonance and mode conversion}

As noted in Section 2, the formulas used for the photon scattering cross sections are applicable only far from vacuum resonance. Therefore, we need to consider the resonance case separately.


The plasma resonance occurs at the plasma density \citep{LaiHo2002}
\begin{equation}\label{11-01}
\rho_V\approx 18.7Y_e^{-1}\left(\frac{E}{1 \text{keV}}\right)^2\left(\frac{B(R)}{10B_\text{QED}}\right)^2f(B)^{-2} \quad \text{g/cm}^3
\end{equation}
where 
$f(B)= (B/5B_\text{QED})^{1/2}$ for $B\gg B_\text{QED}$. If we assume that the maximum radiation is at $\hbar\omega=3kT\sim60$ keV, we obtain $\rho_V\approx 6.7\times10^4(B/10B_\text{QED})$ g/cm$^3$.

When passing through this density photon mode conversion might occur. The conversion is complete if the inequality $\gamma_\text{res}=(E/E_\text{ad})^3\gg1$ is satisfied, where
\begin{equation}\label{11-02}
E_\text{ad}=2.52\left[f(B)\tan\theta\right]^{2/3}\Bigg|1-\frac{\omega_{ci}^2}{\omega^2}\Bigg|^{2/3}\left(\frac{1\text{cm}}{H_\rho}\right)^{1/3} \quad \text{keV}
\end{equation}
here $H_\rho=|\text{d}z/\text{d}\ln\rho|$ is the density scale height along the direction of photon motion, evaluated at $\rho=\rho_V$, and $\hbar\omega_{ci}$ is ion cyclotron energy. 
If $\gamma_\text{res}\ll1$, then the mode conversion is suppressed, and the polarisation of radiation does not change. Therefore, there is always the resonance (the peak in the E mode transparency), and mode conversion occurs only for high-energy photons (with energies $E\gg E_\text{ad}$).



Let us consider the moving plasma. We assume that velocity distribution and mass flow rate are the same as was calculated in Section 3.3. Then the plasma density at the gas sound point $u_s=c_s/v_\text{esc}$, is equal to $\rho_s=\tilde{j}/u_s\sim 10^3$ (for $q_\text{max}\sim 10$), or, converting it to dimensional form, $\rho_b\sim 1$ g/cm$^3$. One sees that the vacuum resonance occurs below the gas sonic point. Then the radiation densities in two modes are equal in the resonance region so that the mode conversion does not play any role there. 


\section{Derivation of the energy equation for O-photons}

In the case under consideration, the optical depth for the O-mode radiation is large, therefore, one can use the diffusion approximation. Namely, the photon distribution function is presented as $n_\text{O}(\omega,\mathbf{k},\mathbf{r})= n_\text{O}^{(0)}(\omega,\mathbf{r})+\delta(\omega,\mathbf{k},\mathbf{r})$;  $\delta(\omega,\mathbf{k},\mathbf{r})\approx(\mathbf{k}\cdot\nabla)n^{(0)}_\text{O}/(\sigma_\text{O}N_e)$, where $n_\text{O}^{(0)}$ is the isotropic part of the distribution and $\delta(\omega,\mathbf{k},\mathbf{r})\ll n_\text{O}^{(0)}(\omega,\mathbf{r})$ \citep{Kaminker1982}. Then the radiation transfer equation is reduced to  \citep{BlandfordPayne1981} 
\begin{equation}\label{d-5-1}
\begin{split}
& (\mathbf{v}_b\cdot\nabla) n_\text{O}^{(0)}-\nabla\cdot\left(D_\text{O}\nabla n_\text{O}^{(0)}\right)-\frac{\omega}{3}\frac{\upartial n_\text{O}^{(0)}}{\upartial\omega}(\nabla\cdot \mathbf{v}_b)= \\
&=\frac{2}{15}\frac{\sigma_T N_e kT_e}{m_ec\omega^2}\frac{\upartial}{\upartial\omega}\omega^4\left[\frac{\upartial n_\text{O}^{(0)}}{\upartial\omega}+\frac{\hbar}{kT_e}(n_\text{O}^{(0)}+(n_\text{O}^{(0)})^2)\right]+\\
&+cN_e\left(\langle\sigma_{\text{E}\rightarrow\text{O}} n_\text{E}\rangle-\langle\sigma_{\text{O}\rightarrow\text{E}}\rangle n_\text{O}^{(0)}\right)
\end{split}
\end{equation}
Here $T_e$ is the electron temperature, $n_\text{E}=n_\text{E}(\omega,\mathbf{k})$ is the distribution function of E-photons, which is considered to be constant because the region is transparent to E-mode. The angle brackets $\langle ...\rangle$ denote averaging over angles, and $D_\text{O}$ is the diffusion tensor of O-photons.
The factor 2/15 before the Comptonization operator takes into account that in the superstrong magnetic field, the Comptonization rate is suppressed because the scattering cross-section contains the factor $\sin^2\theta<1$ and the thermal kinetic energy of electrons is $(1/2)kT_e$, not $(3/2)kT_e$ \citep{BaskoSunyaev75,Lyubarskii88a}. Since we are interested in diffusion along the magnetic field, we only need the longitudinal part of the diffusion tensor,
\begin{equation}\label{d-5-01}
D_\text{O}=c\left\langle \frac{\cos^2\theta}{\sigma_\text{O}(\theta)N_e}\right\rangle=\frac{1}{2\sigma_T N_e}\int\limits_{-1}^{1}\frac{\mu^2\text{d}\mu}{1-[1-(\omega/\omega_{ce})^2]\mu^2}.
\end{equation}
Evaluating integral, we obtain
\begin{equation}\label{d-5-01-1}
D_\text{O}=\frac{\xi c}{\sigma_T N_e}+O\left(\frac{\omega}{\omega_{ce}}\right)
\end{equation}
where 
\begin{equation}\label{xi}
    \xi=\ln\left(2\omega_{ce}/\omega\right)-1.
\end{equation}
The logarithm is a rather slow function therefore $\xi$ is considered as a constant in integrals over $\omega$. We work with spectra that have a peak at $\hbar\omega=3kT\sim 60$ keV, so for simplicity we will take $\xi\sim 5$.

Strictly speaking, the diffusion approximation is applicable only if the optical depth is large in all directions. In our case, the flow is opaque for the O-mode photons at all angles with a possible exception of a small range of angles $\theta<1/\tau^{1/2}$, where $\tau\gg 1$ is the Thomson optical depth. Namely, if \begin{equation}\label{nodiffusion}
    (\omega/\omega_{ce})^2\tau< 1,
\end{equation}  O-photons can escape in this range of angles from a large depth without being scattered, which makes the diffusion approximation invalid. 
In this case, the diffusion operator (the second term in the L.H.S. of equation (\ref{d-5-1})) should be substituted by the general scattering operator
\begin{equation}
    n_{\rm O}^{(0)}(\tau)-\int K(\vert\tau-\tau'\vert) n_{\rm O}^{(0)}(\tau')d\tau',
\end{equation}
where the scattering kernel $K$ is explicitly written by \citet{Lyubarskii88a}.
The eigenvalue of this operator, which is in fact the escape rate of photons, is estimated by \citet{Lyubarskii88b} as $\lambda=\pi^2(\ln 4\tau-2)/4\tau^2$. On the other hand, the eigenvalue of the diffusion operator with the diffusion coefficient  (\ref{d-5-01-1}) is $\lambda=\pi^2\xi/4\tau^2$. This means that in order to take into account the free escape of photons at the condition (\ref{nodiffusion}), one can just substitute the parameter $\xi$ in the form  \begin{equation}\label{xi1}
    \xi=\ln 4\tau-2
\end{equation} 
into the diffusion coefficient (\ref{d-5-01-1}).
It was shown in Section\ 2.2 that the relevant optical depth is $\tau\sim \alpha^{-1}\sim \omega_{ce}/\omega$ therefore, the difference between equations (\ref{xi}) and (\ref{xi1}) is not significant so that the above choice $\xi\sim 5$ may be used in the general case.   


The Comptonization parameter for O-photons is large, $y=(kT_e/m_ec^2)\tau_\text{O}^2\gg 1$, so that the rate of the frequency redistribution exceeds the rate of spatial diffusion. Therefore, the distribution of the O-mode spectrum at each point is close to the Bose-Einstein distribution with the temperature of electrons.

The equation~(\ref{d-5-1}) does not take into account bremsstrahlung processes. 
Both bremsstrahlung and scattering opacities of O-photons along the magnetic field equal to their non-magnetic values, multiplied by $\sin^2\theta$. Therefore, the ratio of the Rosseland mean free-free opacity to the Thomson opacity for O-photons can be written as (\citealt{Potekhin2001})
\begin{equation}\label{free}
\frac{\kappa^\text{O}_{\text{ff}}}{\kappa^\text{O}_\text{es}}\approx\frac{2\times10^4}{c_7}\frac{Z^2}{A}\frac{\rho_b}{T_6^{7/2}}
\end{equation}
where $c_7\approx316.8$
and for simplicity we adopt that the Gaunt factor for free-free transitions $g_\parallel^\text{ff}$ approximately equals to the non-magnetic one. 
At the gas sound point we have $\rho_b<1$ g/cm$^3$ (see Appendix A) and $T\sim10^{8}$ K, 
hence $\kappa^\text{O}_\text{ff}/\kappa^\text{O}_T\approx 10^{-7}Z$. Therefore, free-free transitions can be ignored and Compton scattering dominates the opacity. 

Multiplying the equation~(\ref{d-5-1}) by $\hbar\omega^3/2\pi^2c^3$ and integrating over all frequencies, we obtain the equation for the photon energy density in the O-mode, $\mathcal{E}_\text{O}$, in the form 
 \begin{equation}
 \begin{split}
(\mathbf{v}_b\cdot\nabla) \mathcal{E}_\text{O}-\nabla\cdot\left(D_\text{O}\nabla \mathcal{E}_\text{O}\right)&+\frac{4}{3}\mathcal{E}_\text{O}(\nabla\cdot \mathbf{v}_b)=\\
&=c\sigma_TN_e\alpha^2\left[Q-\left(\frac{T_e}{T}\right)^2\mathcal{E}_\text{O}\right].
\end{split}
\end{equation}
Here $\alpha^2=5(kT/\hbar\omega_{ce})^2$, $T$ is the temperature of the E-photons, which is a parameter in our problem, 
and
\begin{equation}\label{d-5-04}
Q=\frac{1}{\alpha^2\sigma_T}\int\limits_0^{+\infty}\langle\sigma_{\text{E}\rightarrow\text{O}}n_E\rangle\frac{\hbar\omega^3}{2\pi^2c^3}\text{d}\omega
\end{equation}
is the rate of the O-mode energy production due to the conversion of E- to O-modes. In all the integrals,  we used the Wien distribution with the temperature $T_e$ for the O-mode instead of the Bose-Einstein distribution because the maximum of integrands is at $\hbar\omega\sim (3-5) kT_e$, where the difference between two distributions is small.

When integrating the Kompaneets term, we obtain the factor $ (4kT_e-\overline{\hbar\omega})N_e\mathcal{E}_\text{O}$, where $\overline{\hbar\omega}$ is the energy-weighted mean photon frequency. This term vanishes because the specific heat of the plasma is small with the specific heat of radiation. Then $kT_e=\overline{\hbar\omega}/4$. 


The radiation from the fireball is thermal with the temperature $T$; therefore $Q\propto T^4$. For the Wien distribution, one can write $\mathcal{E}_\text{O}\sim T_e^4\exp(-\zeta/kT_e)$, where $\zeta>0$ is the chemical potential of photons. At low altitudes, where plasma density is high enough, we can assume that $\zeta\lesssim kT_e$, therefore we could substitute $(T_e/T)^2\sim (\mathcal{E}_\text{O}/Q)^{1/2}$. At higher altitudes, where $\zeta$ can be large (which means that $\mathcal{E}_\text{O}$ is small), such a substitution would be incorrect. 
However, the plasma density is also low there, 
so the R.H.S. of the energy equation is small. Therefore we do not expect a significant error if we write the energy equation in the form: 
\begin{equation}\label{d-5-03}
\begin{split}
(\mathbf{v}_b\cdot\nabla) \mathcal{E}_\text{O}-\nabla\cdot\left(D_\text{O}\nabla \mathcal{E}_\text{O}\right)&+\frac{4}{3}\mathcal{E}_\text{O}(\nabla\cdot \mathbf{v}_b) = \\
&=c\sigma_TN_e\alpha^2\left(Q-\frac{\mathcal{E}_\text{O}^{3/2}}{Q^{1/2}}\right).
\end{split}
\end{equation}

In the static case,   $\mathbf{v}_b=0$, the equation~(\ref{d-5-03}) is reduced to the ordinary diffusion equation. Without the adiabatic cooling the radiation temperature remains constant; varies only the chemical potential, i.e. the photon density.  Therefore, in this case, $T_e=T$, so that the energy equation may be written as  
 \begin{equation}\label{d-5-03-1}
-\nabla\cdot\left(D_\text{O}\nabla \mathcal{E}_\text{O}\right)=c\sigma_TN_e\alpha^2\left(Q-\mathcal{E}_\text{O}\right).
\end{equation}

\section{Dipole coordinate system}

Let us define the  dipole coordinates $(\mu,\chi,\phi)$ via the spherical coordinates $(r,\theta, \phi)$  according to the formulas
\begin{gather}\label{d-01}
\mu=-\frac{\cos\theta}{r^2}, \quad \chi=\frac{\sin^2\theta}{r}, \quad \phi=\phi.
\end{gather}
The basis vectors of the new coordinate system are
\begin{gather}\label{d-02}
\mathbf{e}_i=\frac{1}{h_i}\frac{\partial\mathbf{r}}{\partial q_i}
\end{gather}
where the index $i$ has the values $(\mu,\chi,\phi)$, and $h_i=|\partial\mathbf{r}/\partial q_i|$ -- the Lame coefficient of the $i-$th coordinate.
In the spherical coordinate system we have $\mathbf{r}=r\mathbf{e}_r$, therefore 
\begin{gather}\label{d-03}
\frac{\partial\mathbf{r}}{\partial \mu}=\frac{\partial r}{\partial \mu}\mathbf{e}_r+r\frac{\partial\mathbf{e}_r}{\partial\mu}=\frac{\partial r}{\partial \mu}\mathbf{e}_r+r\frac{\partial\mathbf{e}_r}{\partial \theta}\frac{\partial\theta}{\partial \mu}.
\end{gather}
It follows from equations \eqref{d-01} that $\mu^2r^4+\chi r=1$. Differentiating this equality with respect to $\mu$ at constant $\chi$, we get  $\partial r/\partial \mu$. On the other hand, by eliminating $r$, equations \eqref{d-01} can be reduced to $\chi^2\cos\theta+\mu\sin^4\theta=0$. Differentiating this relation with respect to $\mu$ yields $\partial \theta /\partial\mu$. Then we find
\begin{gather}\label{d-04}
\frac{\partial\mathbf{r}}{\partial \mu}=\frac{r^3}{\delta}\left(2\cos\theta\,\mathbf{e}_r+\sin\theta\,\mathbf{e}_\theta\right)
\end{gather}
where $\delta = \sqrt{1+3\cos^2\theta}$. Now the unit vector $\mathbf{e}_\mu$ and the corresponding Lame coefficient are found as
\begin{gather}\label{d-05}
\mathbf{e}_\mu=\frac{2\cos\theta}{\delta}\mathbf{e}_r+\frac{\sin\theta}{\delta}\mathbf{e}_\theta, \quad h_\mu=\frac{r^3}{\delta}
\end{gather}
Similar calculations lead to the formulas 
\begin{gather}\label{d-06}
\mathbf{e}_\chi=-\frac{\sin\theta}{\delta}\mathbf{e}_r+\frac{2\cos\theta}{\delta}\mathbf{e}_\theta, \quad h_\chi=\frac{r^2}{\delta\sin\theta};
\\
\label{d-07}
\mathbf{e}_\phi=\mathbf{e}_\phi, \quad h_\phi=r\sin\theta.
\end{gather}
It is easy to check that the system of the obtained unit vectors forms an orthonormal basis,  $\mathbf{e}_i\cdot\mathbf{e}_j=\delta_{ij}$. Moreover, unit vectors $(\mathbf{e}_\mu,\mathbf{e}_\chi,\mathbf{e}_\phi)$ forms a right-handed basis. Indeed, one can check that
\begin{equation}\label{d-07-1}
\mathbf{e}_\mu\times\mathbf{e}_\chi=\mathbf{e}_\phi, \quad \mathbf{e}_\chi\times\mathbf{e}_\phi=\mathbf{e}_\mu, \quad \mathbf{e}_\phi\times\mathbf{e}_\mu=\mathbf{e}_\chi.
\end{equation}
Finally, as an example, we write the gradient and the divergence differential operators in the new coordinate system. The gradient becomes 
\begin{gather}\label{d-08}
\nabla A = \frac{\delta}{r^3}\frac{\upartial A}{\upartial \mu}\mathbf{e}_\mu+\frac{\delta \sin\theta}{r^2}\frac{\upartial A}{\upartial \chi}\mathbf{e}_\chi+\frac{1}{r\sin\theta}\frac{\upartial A}{\upartial\phi}\mathbf{e}_\phi
\end{gather}
the divergence
\begin{equation}\label{d-09}
\begin{split}
\nabla\cdot\mathbf{A}&=\frac{\delta ^2}{r^6}\bigg[\frac{\partial}{\partial\mu}\left(\frac{r^3}{\delta}A_\mu\right)+\frac{\upartial}{\upartial\theta}\left(\frac{r^4\sin\theta}{\delta}A_\chi\right)+ \\
&+\frac{\upartial}{\upartial\phi}\left(\frac{r^5}{\delta^2\sin\theta}A_\phi\right)\bigg].
\end{split}
\end{equation}
The exact form for other differential operators can be obtained from well-known formulas.


\bsp	
\label{lastpage}
\end{document}